\documentclass{iopart}
\usepackage[dvips]{graphicx}
\usepackage{bm}

\newcommand{\be}{\begin{equation}}
\newcommand{\ee}{\end{equation}}
\newcommand{\bea}{\begin{eqnarray}}
\newcommand{\eea}{\end{eqnarray}}
\newcommand{\nn}{\nonumber}
\newcommand*{\MEU}{Laboratoire Univers et Th\'eories, UMR 8102
  du C.N.R.S., Observatoire de Paris, F-92195 Meudon Cedex, France}
\newcommand*{\IAA}{Instituto de Astrof\'{\i}sica de Andaluc\'{\i}a,
  CSIC, Apartado Postal 3004, Granada 
        18080, Spain} 
 
\begin{document}

\title{A fast stroboscopic spectral method for rotating systems in numerical relativity }

\author{Silvano Bonazzola$^{1}$\footnote{Email address:
    Silvano.Bonazzola@obspm.fr},
Jos\'e Luis Jaramillo$^{2}$\footnote{Email address: jarama@iaa.es} 
and J\'{e}r\^{o}me Novak$^1$\footnote{Email address: Jerome.Novak@obspm.fr} }
\address{$^1$ \MEU}
\address{$^2$ \IAA}

\date{May, 22nd 2007}

\begin{abstract}
  We present a numerical technique for solving evolution equations, as the
  wave equation, in the description of rotating astrophysical compact objects
  in comoving coordinates, which avoids the problems associated with the light
  cylinder. The technique implements a fast spectral matching between two
  domains in relative rotation: an inner spherical domain, comoving with the
  sources and lying strictly inside the light cylinder, and an outer inertial
  spherical shell.  Even though the emphasis is placed on spectral techniques,
  the matching is independent of the specific manner in which equations are
  solved inside each domain, and can be adapted to different schemes. We
  illustrate the strategy with some simple but representative examples.

\end{abstract}


\maketitle

\section{Introduction}

A general motivation for undertaking this project is suggested by the
numerical study of astrophysical compact objects, e.g. neutron stars and
black holes, individually rotating or in bound binary systems,
where the dynamics of the interaction field is described by a
system of equations which is (at least, in part) hyperbolic. An example of
this is provided by Einstein equations in General Relativity or the Maxwell
equations in modeling the pulsar magnetosphere. The main objective of this
paper is to present a general numerical technique for solving the dynamics of
such systems, with the example of the scalar wave equation as the equation
governing the evolution of the system.

The study of the general motion of compact finite bodies, say black holes (BH)
or neutron stars (NS) for concreteness, in a fixed numerical grid represents a
serious challenge from a numerical point of view. Examples of this Eulerian
approach to the motion problem can be found in Refs.~\cite{AlcubS92,BonaMS95}.
Recent examples of moving BH in the context of binary systems have been
successfully developed an implemented both using excision techniques
\cite{Preto05,SperhKLal05,SzilaPRal06} and moving punctures
\cite{CampaLMal06,BakerCCal06, BruegGHal06}. However, we would find
unfortunate if these extraordinary achievements actually shadowed the
development of parallel approaches which employ independent analytical
formulations and/or numerical methods. This is the case of  constrained
evolution formalisms (e.g. see Ref. \cite{BonazGGN04}) and of the use of
spectral methods (already initiated in the binary BH case in Ref.
\cite{ScheePLKRT06}). More concretely, the use of spectral methods seem to
favor an excision approach for BH evolution where the coordinate location of
the excised surface do not change significantly in time \cite{ScheePLKRT06}.
On the other hand constrained evolutions schemes, in combination with a
comoving coordinate systems, can be employed to endow the excised surface with
a geometrical character by enforcing appropriate inner boundary conditions in
some of the elliptic (constraint) equations, something that can be relevant
both for the numerical stability of the code and the study of some
quasi-local geometrical properties of BH space-times \cite{trap_hor}. These
elements indicate the interest in pursuing the study of corotating coordinate
systems.

Regarding the presence of matter, namely the NS case, experience in our group
shows the convenience of using a Lagrangian scheme to describe the surface of
the NS and a Eulerian one for the interior. Details of this technique can be
found in Refs.~\cite{Gourg91} and in \cite{BonazGM98} where a steady state
approximation was used. Building accurate non-stationary numerical models of
magnetized rotating neutron stars and of the surrounding magnetosphere might
thus also be easier with such codes, able to naturally follow the surface of the
star. Therefore, an interesting alternative strategy to the Eulerian approach consists
in reducing the motion of the body in the grid
by choosing a comoving coordinate system.
However, the use of a single rotating coordinate system faces the problems
coming from the superluminal motion of the grid at distances to the rotation
center larger than the light cylinder radius (see next section for the
definition of the latter).  Part of these problems, related to the
tensorial nature of the rotating fields, can be handled by using
appropriate techniques (e.g. the ``dual-coordinate frames'' introduced in Ref.
\cite{ScheePLKRT06}).

The aim of this paper is to propose a complementary approach that bypasses the
light cylinder issue in the case of a scalar field wave equation, using an
intermediate strategy in which space is split into two domains: first, an
internal spherical one of radius $R_*$ containing the sources and set inside
the light cylinder, and then an external domain with inner boundary given by
the sphere $r=R_*$ and whose outer radius is larger than the considered
wavelength in order to impose approximate (or exact) outgoing wave conditions.
The exterior domain is described in terms of non-rotating coordinates and
simultaneously, a rigidly\footnote{The notion of ``rigidity'' employed here,
  is defined in terms of a fiducial time-independent flat 3-metric $f_{ij}$,
  as introduced in the constrained evolution formalism presented in Ref.
  \cite{BonazGGN04}.} rotating coordinate system is set in the interior
domain whose angular velocity is fixed by a physical scale in the problem
(e.g.  the orbital frequency in a binary problem). Each domain could be
further decomposed into smaller ones, but this paper focuses on the
matching conditions and not in the specific manner in which the equations are
solved inside each domain.  The main problem we address here is then the
matching technique between domains in relative rotation in a multi-domain
spectral scheme. In particular, if we consider the spherical
boundary between the inner and outer domains, and denote by $N_\theta$ and
$N_\varphi$ the number of sampling points in the $\theta$ and $\varphi$
spherical-coordinate directions, then the technique for matching the solutions
at this intermediate boundary is said to be fast in the following sense. The
most straightforward approach is just to write the matching system, using a
spectral summation, at every point of each grid, which requires $N_\theta^2
N_\phi^2 $ operations. Our approach here can be performed with only $ N_\theta
\cdot N_\varphi $ instead, with a so-called {\it stroboscopic} technique based
on spectral methods.

The paper is organized as follows: Section~\ref{s:model} describes the simple
model we want to study and specifies the light cylinder problem,
Section~\ref{numerical_strategy} gives the numerical details of our matching
technique and some outlines of the implementation, Section~\ref{s:tests} shows
the results of two test-problems, and finally Section~\ref{s:conc}
contains concluding remarks.  We emphasize that in this work we address the
description of a technique enabling us to study the dynamics of a system with
a rotating source.  Even though the main problem we have in mind is the study
of the evolution problem for some given initial data, we can also employ these
resolution techniques in order to construct the initial data themselves by
assuming additional hypotheses and/or approximations for some given terms in the
equations.  In particular, in the description of the present technique we do
not make any assumption about the existence of approximate Killing vectors,
something that would have to be incorporated in the equations themselves in
the construction of initial data of binaries in quasi-circular motion (see
Refs.~\cite{Detwe94,BonazGM97,FriedUS02}) for different approaches, and
below).

\section{Light cylinder and simplified mathematical models}
\label{s:model}

The resolution of the full Einstein Equations (EE) can be reduced to solve a
couple of wave equations for two independent scalar potentials if one uses a
constrained scheme, where the ten EE have been decomposed as four elliptic
constraint equations, four gauge degrees of freedom and two scalar wave-like
evolution equations~\cite{BonazGGN04}. In this section we consider the
linearized version of the actual problem for one of the potentials. This is
quite general since in the overall problem for the solution of the EE, we
shall perform exactly the same matching for every wave-like equation.

In particular, consider a non-rotating coordinate system $(t, r , \theta,
\varphi)$ and a field $\Phi$ satisfying 
\be
\label{e:wave_equation}
\left(\frac{1}{c^2}\partial^2_t - \Delta\right)\Phi = n, 
\ee
where $c$ is the propagation speed of the field and $n$ denotes the matter
mass density (renormalized with a $4\pi G$ factor). We assume that matter is
contained in a compact region bounded by characteristic radius $R_*$.
Defining a spherical coordinate system rotating rigidly with constant 
angular velocity $\Omega$, $(t'=t, r'=r ,
\theta'=\theta, \varphi'=\varphi+\Omega\; t)$, and writing explicitly the
Laplacian, the wave equation becomes
\bea
\label{e:rotating_wave_equation}
\left[\frac{1}{c^2}\partial^2_{t'}+2 \frac{\Omega}{c^2}\partial_{t'}\partial_{\varphi'}
-\partial^2_{r'} - \frac{2}{r'}\partial_{r'}\right.  \\
\left. -\frac{1}{r'^2}\left(\partial^2_{\theta'} 
+\frac{\mathrm{cos}{\theta'}}{\mathrm{sin}{\theta'}}
\partial_{\theta'} + 
\frac{1}{\mathrm{sin}^2{\theta'}}\left(1 - \left(\frac
 {\Omega r' \mathrm{sin}{\theta'}}{c}\right)^2 \right)
\partial^2_{\varphi'}\right)\right]\Phi = n \nn
\eea
We note that in our formulation of the problem, there appears a second
characteristic scale, given by the vanishing of the factor $\left(1 -
  \left(\frac{\Omega r \mathrm{sin}\theta'}{c}\right)^2\right)$.  This defines
the {\it light cylinder} of radius $R_L = \frac{c}{\Omega}$, with respect
to the rotation axis, where the vector
$\partial_{t'}$ (corresponding to $\partial_t-\Omega \partial_\varphi$ in the
inertial system) becomes null. The total operator in
Eq.~(\ref{e:rotating_wave_equation}) is hyperbolic everywhere. However we note 
that, even though the spatial part of the operator is negative definite in the
region inside the light cylinder, $r<R_L$, this feature is lost for $r>R_L$.
Motivated by this remark (see discussion below), we will always assume the
hypothesis that $R_*<R_L$.  This assumption will be eventually justified on
physical grounds.

As mentioned above, in the present work we study the dynamical evolution of a
system with a given matter source $n(t,r,\theta,\varphi)$. Then, our strategy
to describe the dynamics consists in solving the equations in different
coordinate systems depending on the domain.  In concrete terms: a) we choose
rotating coordinates in the domain containing matter, i.e.  we solve
Eq.~(\ref{e:rotating_wave_equation}) for $r \le R_*$, and b) we employ an
inertial coordinate system in the exterior region, i.e for $ r \ge R_*$ we
solve Eq.~(\ref{e:wave_equation}).

Even though we are focused on the general evolution problem, it is convenient
to highlight that the existence of the light cylinder is an important issue in
the construction of 
quasi-stationary rotating configurations of binary systems.
The introduction of a helical Killing vector has been used in the literature
(cf. \cite{Detwe94,BonazGM97,FriedUS02}) in order to model slow-motion
adiabatic configurations of binary NS and BH systems.  In particular,
regarding Eq.~(\ref{e:rotating_wave_equation}) this imposition of the vector
$\partial_{t'}$ as a helical symmetry of the solution, implies the vanishing
of the first two terms in the left-hand-side of the equation.  A discussion of
the numerical technical issues of the resulting mixed-type partial
differential equation, that is elliptic inside the light cylinder and
hyperbolic outside, can be found in the series of works on quasi-stationary binary
inspiral \cite{WhelaR99,WhelaKP00,WhelaBLP02,Price04}, as well
as the so-called ``periodic standing-wave'' approximation
\cite{Andraal04, BromlOP05,BeetlBP06,LauP07}.
See also Ref. \cite{Torre03} for a study of its mathematical well-posedness.
In this context of potential numerical problems associated with the type of
the spatial part of the differential operator, the advantage of our approach
is that the light cylinder problem is completely avoided, since $R_* < R_L $.
Indeed, in the interior domain the differential operator presents a standard
``negative'' spatial part and there are no problems in solving numerically
Eq.~(\ref{e:rotating_wave_equation}), whereas no light cylinder problem shows
up in Eq.~(\ref{e:wave_equation}) in the exterior domain.

We conclude the section by emphasizing that with this point of view, the light
cylinder problem is replaced by a matching problem, which is much easier to
treat numerically. We therefore state the main problem to be addressed in this
approach, and which defines the goal of the paper: the implementation of the
matching between the solutions in both domains. Next section presents a
fast\footnote{ By fast we mean that the number of operations is not larger
  than $ N \cdot \mathrm{log} N $ for each dimension, where $N$ denotes the
  number of degrees of freedom in a given dimension. As it has been said, in
  the present case the number of operations is proportional to $N_\theta \cdot
  N_\varphi$.}  matching numerical algorithm.

\section{Numerical strategy}
\label{numerical_strategy}

We introduce the following dimensionless variables $\xi$ and $\tau$ in the
inner and outer domains:
\bea
\label{xi_tau}
\xi'&=&\frac{r'}{R_*}, \; \; \; \tau'=\frac{t'}{P} 
\quad (\hbox{for } r'\le R_*), \quad \nn
\hbox{ and } \\
\xi&=&\frac{r}{R_*}, \; \; \; \tau=\frac{t}{P} 
\quad (\hbox{for } r\ge R_*) ,
\eea
where $P=2 \pi/ \Omega$ is the rotation period.  Making use of these
coordinates and denoting the field $\Phi$ by $\Phi^<$ in the interior domain
($0\leq\xi\leq 1$), and by $\Phi^> $ in the exterior domain ($\xi\geq 1$), we
write Eqs.~(\ref{e:wave_equation}) and (\ref{e:rotating_wave_equation}) in the
following dimensionless form
\bea
&&\label{e:ad_wave_equation}
\left (\frac{\partial ^2}{\partial \tau^2}-C^2 \bar{\Delta} \right)
\Phi^>=0 \ \ ,  \\
&&\label{e:ad_rotating_wave equation} 
\left[ \frac{\partial^2}{\partial \tau'^2}+2(2\pi) \frac{\partial^2}
{\partial \tau' \partial \phi'} +(2\pi)^2 \frac{\partial^2}{\partial \phi'^2}
-C^2 \bar{\Delta} \right ] \Phi^<=\bar{n}
\eea
where 
\be\label{ad_C_light}
C=2\pi \frac{R_L}{R_*},
\ee
is the dimensionless ``light velocity'', $\bar{\Delta}$ is the Laplacian in
terms of the dimensionless radial coordinate $\xi$, and $\bar{n}=R_*^2 C^2
\cdot n$ is the dimensionless matter density
(note that the dimensionless angular velocity is $2 \pi$).\\
In order to simplify the presentation, and since the notation should be
self-explanatory in the following, we shall drop the ``prime'' for the
equations in the rotating grid hereafter.

\subsection{Stroboscopic matching}

We look for a solution of the above Eqs.~(\ref{e:ad_wave_equation}) and
(\ref{e:ad_rotating_wave equation}) by expanding the fields $\Phi$ and
$\bar{n}$ in spherical harmonics
\bea
\Phi= \sum_{\ell m} P^m_\ell(\cos \theta) \left[  a_{\ell m} \cos (m \varphi) +b_{\ell m}
 \sin (m \varphi) \right] ,\\ 
\bar{n} = \sum_{\ell m} P^m_\ell(\cos \theta) \left[  \bar{n}^c_{\ell m} \cos
  (m \varphi) +\bar{n}^s_{\ell m}  \sin (m \varphi) \right]. 
\eea
where the coefficients $a_{\ell m}(\xi,\tau)$ and $b_{\ell m}(\xi,\tau)$ are
the unknown functions, and $P^m_\ell(\theta)$ are the Legendre associate
functions.  Similarly, $\bar{n}^c_{\ell m}$ and $\bar{n}^s_{\ell m} $ are the
components of the source $\bar{n}$. In the inner domain $(\xi\le 1)$, we
obtain a system of two coupled partial differential equations for each value
of $\ell$ and $m$ (recall that we now use $\tau = \tau'$):  
\begin{eqnarray}
&& \bigg[ \frac{\partial^2}{\partial \tau^2}                                
-C^2 \left( \frac{\partial^2}{\partial \xi^2} +\frac{2}{\xi} 
\frac{\partial}{\partial \xi} -\frac{\ell(\ell+1)}{\xi^2} \right)      \nonumber \\ 
&&-(2\pi)^2 m^2 \bigg]\,  a^<_{\ell m} +2(2\pi) m
\frac{\partial}{\partial \tau} b^<_{\ell m}  
=\bar{n}^c_{\ell m}, \label{e:rotating_harm_cos_wave_equation}  
\end{eqnarray}
and
\begin{eqnarray}
&&\bigg[ \frac{\partial^2}{\partial \tau^2} 
-C^2 \left( \frac{\partial^2}{\partial \xi^2} +\frac{2}{\xi}
\frac{\partial}{\partial \xi} -\frac{\ell(\ell+1)}{\xi^2} \right)    \nonumber  \\ 
&&-(2\pi)^2 m^2\bigg] \,  b^<_{\ell m} -2(2 \pi) m 
\frac{\partial}{\partial \tau}\,a^<_{\ell m}
 =\bar{n}^s_{\ell m}. 
\label{e:rotating_harm_sin_wave_equation}
\end{eqnarray}

Similarly, for the external domain $(\xi\ge 1)$ we find:
\bea\label{e:galil_harm_cos_wave_equation} 
\left [ \frac{\partial^2}{\partial \tau^2} -C^2 \left(
 \frac{\partial^2}{\partial \xi^2} + \frac{2}{\xi} \frac{\partial}{\partial \xi} 
-\frac{\ell(\ell+1)}{\xi^2} \right) \right] a^>_{\ell m}&=&0 \\
\label{e:galil_harm_sin_wave_equation}
\left [\frac{\partial^2}{\partial \tau^2} -C^2 \left(
\frac{\partial ^2}{\partial \xi^2} +\frac{2}{\xi} \frac{\partial}
 {\partial \xi} -\frac{\ell(\ell+1)}{\xi^2} \right ) \right ] b^>_{\ell m}&=&0.
\eea
By taking into account the expression of the ``light velocity'' given by
Eq.~(\ref{ad_C_light}) and the inequality $m \leq \ell$, we can see that the
second-order differential spatial operator appearing in
Eqs.~(\ref{e:rotating_harm_cos_wave_equation}),
(\ref{e:rotating_harm_sin_wave_equation}) is of definite type if $R_L \ge
R_*$.

In order to solve these equations we make use of a multi-domain spectral
technique in the radial direction and finite differences in the time variable.
In our specific implementation, for each couple of values $(\ell, m)$ and in
each domain we construct a particular solution as well as the relevant
homogeneous solutions; in our simple case, regularity of the solution at the
origin leaves a single homogeneous solution in the interior domain
\cite{BM90}, whereas in the exterior domain there are two homogeneous
solutions.  Solutions to
Eqs.~(\ref{e:rotating_harm_cos_wave_equation})-(\ref{e:rotating_harm_sin_wave_equation})
in the inner domain and to
(\ref{e:galil_harm_cos_wave_equation})-(\ref{e:galil_harm_sin_wave_equation})
in the outer domain are written as linear combinations of the calculated
homogeneous and particular solutions, thus involving in total three
coefficients to be fixed. The latter are determined by: {\it i\/}) imposing an
outgoing Sommerfeld wave condition on $\Phi^>$ at the outer boundary of the
exterior domain (see e.g. Ref. \cite{NovakB}), 
\be\label{Somme} 
\left[ \bigg( \frac{\partial}{\partial \tau}+\frac{\partial }{\partial
    \xi}+\frac{1}{\xi} \bigg) \Phi^> \right]_{\xi=\xi_D}=0,
\ee 
where $\xi_D$ is the external radius of the outer domain; {\it ii\/}) enforcing the
continuity of the fields $\Phi^<$ and $\Phi^>$, and {\it iii\/}) of their radial
derivatives $\Phi'^<$ and $\Phi'^>$, at the intermediate boundary $\xi=1$.
This makes three conditions for three homogeneous solutions and the matching
system is invertible.

The following discussions focus on the two last points, i.e. the matching
conditions at $\xi=1$, for a given value of the angle $\theta_i$, $i \in \{1,
..., N_\theta\}$.  In our scheme we employ the same number of sampling points
in the $\varphi$ direction, $N_\varphi$, in the two grids.  Using an index
$j\in\{1,..., N_\varphi\}$, let us denote by $\Phi^<_j(\tau)$ and
${\Phi'}^<_j(\tau)$ the values of the field $\Phi^<(\tau, \xi=1,
\theta=\theta_i, \varphi=\varphi_j)$ and its radial derivative $\Phi'^<(\tau,
\xi=1, \theta=\theta_i, \varphi=\varphi_j)$ on the $\varphi$-sampling points
in the inner grid, and by $\Phi^>_j(\tau)$ and ${\Phi'}^>_j(\tau)$
the analogous values of $\Phi^>(\tau, \xi=1, \theta=\theta_i,
\varphi=\varphi_j)$ and $\Phi'^>(\tau, \xi=1, \theta=\theta_i,
\varphi=\varphi_j)$ in the outer domain.

\begin{figure}
  \centerline{\includegraphics[scale=0.4]{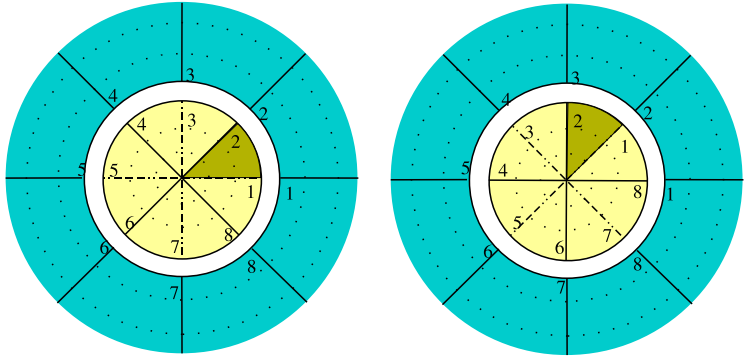}}
  \caption[]{
    \label{f:Figure1}
    Choice of evolution step $\delta \tau$, adjusted to the sampling interval
    $\delta \varphi$, so as to guarantee the ({\it stroboscopic}) coincidence
    of the $\varphi-$sampling points corresponding to the interior and the
    exterior grids along the whole time evolution.}
\end{figure}

\begin{figure}
  \centerline{\includegraphics[scale=0.4]{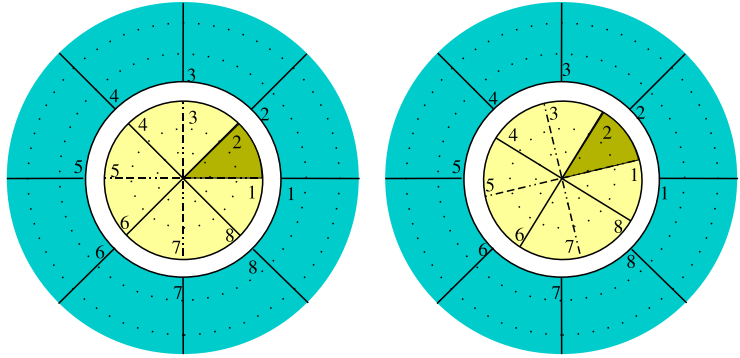}}
  \caption[]{
    \label{f:Figure2}  
    Choice of evolution step $\delta \tau$ as an integer fraction (in this
    case $K=3$) of the $\varphi-$sampling interval $\delta \varphi$. In this
    case, interpolation of the points is needed for enforcing the matching.
    This is achieved in a {\it fast} manner by employing a spectral
    oversampling technique based on the Gauss-Radau points.}
\end{figure}

Then, at the initial time $\tau=\tau_\mathrm{o} $, we choose the sampling
points of the two grids to coincide.  In this situation we use strong matching
techniques\footnote{Alternatively, variational (weak) matching methods could be used.
  Penalty terms can also be introduced if this is suggested by the complexity
  of the system.}, that are simply expressed as: 
\be
\label{e:matching}
\begin{array}{ccc}
\Phi^<_1(\tau_0) = \Phi^>_1(\tau_0) \ \ & , & \ \ \Phi'^<_1(\tau_0) = \Phi'^>_1(\tau_0) \nn \\
\Phi^<_2(\tau_0)= \Phi^>_2(\tau_0) \ \ & , & \ \ \Phi'^<_2(\tau_0) = \Phi'^>_2(\tau_0) \nn \\
&... & \\
\Phi^<_{N_\varphi}(\tau_0) = \Phi^>_{N_\varphi}(\tau_0) \ \ & , & \ \ \Phi'^<_{N_\varphi}(\tau_0) = \Phi'^>_{N_\varphi}(\tau_0) \nn
\end{array}
\ee
This makes $2N_\varphi$ conditions for each value $\theta_i$, which is the
right number of matching relations.
Regarding the choice of the evolution time step $\delta \tau$, two situations
may occur: if one uses an explicit time-evolution scheme, it must satisfy the
stability Courant conditions; on the other hand, if the time-integration
scheme is implicit (more precisely, $A$-stable), then the time-step is not
limited. Let us first suppose that this is the case and let us choose $\delta
\tau= \frac{\delta \varphi}{2\pi}$, where $\delta
\varphi=\frac{2\pi}{N_\varphi}$ is the sampling $\varphi-$angular interval.
With this choice of $\delta \tau$ the sampling points at $\tau=\tau_\mathrm{o}
+k\cdot \delta \tau$, with $k\in \{0,1,2,...\}$, do coincide again at any later
time (see Fig. \ref{f:Figure1}).
The matching for a step $k$ is simply given by:
\be
\label{e:rotated_matching}
\begin{array}{ccc}
\Phi^<_1 = \Phi^>_{(1+k)\mathrm{mod}(N_\varphi)} \ \ & , & \ \ 
\Phi'^<_1 = \Phi'^>_{(1+k)\mathrm{mod}(N_\varphi)} \nn \\
\Phi^<_2 = \Phi^>_{(2+k)\mathrm{mod}(N_\varphi)} \ \ & , & \ \ 
\Phi'^<_2 = \Phi'^>_{(2+k)\mathrm{mod}(N_\varphi)}   \\
&... & \\
\Phi^<_{N_\varphi} = \Phi^>_{(N_\varphi+k)\mathrm{mod}(N_\varphi)} \ \ & , & \ \ \Phi'^<_ {N_\varphi}= \Phi'^>_{(N_\varphi+k)\mathrm{mod}(N_\varphi)}. \nn
\end{array}
\ee 
We now consider the more general
case in which $\delta \tau$ is bound to be
smaller than a given critical value, given for instance by the Courant
condition.  Our strategy consists in choosing a $\delta \tau$, satisfying
$\delta \tau =\frac{1}{2\pi\cdot K} \delta \varphi$ where $K$ is a
sufficiently large integer number. In this case, the sampling points only
coincide after $K$ time steps, and therefore we need to interpolate at 
supplementary points in the $\varphi$ coordinate (see Figure \ref{f:Figure2}).
At this stage, spectral techniques provide an straightforward manner to
proceed by performing an ``oversampling'' of the spectral description. That
is, given the original $N_\varphi$ coefficients codifying the behavior of the
function in the $\varphi$ direction and corresponding to the $N_\varphi$
sampling points [e.g.  $(a_{\ell1},a_{\ell2},..., a_{\ell N_{\varphi}}$) for a
given $\ell$], one increases the number of sampling points in $\varphi$ by a
factor $K$, in such a manner that the new coefficients, counterpart of the new
points, are set to zero: the information is encoded in the same number of
non-vanishing coefficients but in a space with a bigger number of degrees of
freedom:
\be
\label{e:oversampling}
(a_{\ell1},a_{\ell2},..., a_{\ell N_{\varphi}},
\underbrace{\overbrace{0,...,0}^{N_\varphi \ \hbox{times}},...,
  0,...,0}_{(K-1)\cdot N_\varphi \ \hbox{times}}).  \ee The values at the new
(interpolation) points can be obtained by performing the inverse Fourier
transform on the larger set of coefficients.  Moreover, a Fast Fourier
Transform algorithm could be used only involving a number of operations
proportional to $(N_\varphi\cdot K)\mathrm{log}(N_\varphi\cdot K)$ for each
$\ell$.  Once the matching is done, in an analogous manner to the one
described in conditions (\ref{e:rotated_matching}), only $N_\varphi$
coefficients are kept in order to codify the function $\Phi$ at each domain.
Proceeding in this way, the functions $\Phi^<$, $\Phi'^<$ and $\Phi^>$,
$\Phi'^>$ can be interpolated at the Gauss-Radau sampling points of the
Fourier expansion. These Gauss-Radau sampling points are ``smart'' sampling
points for which the following identity holds
$$ \int_0^{2\pi}\cos(k \phi) \cos(m\phi) d\phi= \sum_{j=0}^{N-1}
  \cos(2\pi k j/N_\varphi) \cos(2\pi m j/N_\varphi) 
$$ 
$k, \;m, \; j$ being integer positive numbers and $ k, \, m \le N_\varphi/2$
(analogous relations for other trigonometric functions).  If the solution of
the above system of equations does not contain spatial frequencies (in the
Fourier expansion) larger than $N_\varphi/2$, then a solution obtained with a
number of degrees of freedom higher than $N_\varphi$ will give the same result
(within round-off errors); therefore the interpolation is exact.
 
Even though the previous discussion illustrates the underlying philosophy, the
method presented above is not the most efficient one.  As a matter of fact,
there is a manner to by-pass the inverse Fourier transform step, gaining a
factor $\log N$ in each angular direction. This is accomplished by making the
matching in the coefficient space rather than in the configuration values of
the functions $\Phi$. For this, we must expand the inner and outer solutions
in the same $\varphi$-coordinate expansion bases. This can be achieved
straightforwardly, since the respective bases [$\mathrm{cos}(m\varphi')$ and
$\mathrm{sin}(m\varphi')$ in the inner domain, and $\mathrm{cos}(m\varphi)$ and
$\mathrm{sin}(m\varphi)$ in the outer domain], are related by a simple
rotation: $\delta \varphi = \varphi' -\varphi= \Omega \delta t =
2\pi \delta\tau$.  Therefore,
given the function $F(\varphi') = \Phi^< (\tau, \xi=1, \theta, \varphi')$ in
the inner domain with \be F(\varphi') = \sum_{m=0} ^{N_\varphi-1} \left ( a'_m
  \cos(m \varphi')+ b'_m \sin(m\varphi') \right ), \ee the standard
trigonometric relations provide the coefficients $a_m$ and $b_m$ in the
$\varphi$ bases employed in the outer domain, and necessary for performing the
matching: 
\bea
a_m&=&a'_m \cos(m \, \delta \varphi) +b'_m \sin(m \, \delta \varphi) \nn  \\
b_m&=&-a'_m \sin(m \, \delta \varphi)+b'_m \cos(m \, \delta \varphi) 
\eea 
If $\delta \varphi =2 \pi/N_\varphi $, this is equivalent to sample the
function $F(\varphi)$ at the Gauss-Radau points and to come back in the
Fourier space.  In other words, the matching is done without leaving the
Fourier space, but at the cost of (again) oversampling points in the
$\varphi$-direction by a factor $K$. The number of operation is proportional
to $N_\theta \cdot N_\varphi$.
A bonus of this strategy, that we refer to as {\it stroboscopic}, is the
following: the technique described above prevents the appearance of spurious
time frequency terms generated by the beating between the rotation frequency
and the time sampling frequency. This terms could spoil the study of the
numerically constructed signal, in e.g. an a posteriori time Fourier transform
analysis of the emitted radiation.

\subsection{Implementation}

Equations~(\ref{e:rotating_harm_cos_wave_equation})-(\ref{e:rotating_harm_sin_wave_equation})
and
(\ref{e:galil_harm_cos_wave_equation})-(\ref{e:galil_harm_sin_wave_equation})
are solved by using the spectral methods developed in our group
\cite{Meudon}.

For the time evolution in the non-rotating domain [i.e. Eqs.~(\ref{e:galil_harm_cos_wave_equation}) and
(\ref{e:galil_harm_sin_wave_equation})] we use a second-order implicit
Crank-Nicolson scheme.  In addition, Sommerfeld outgoing boundary conditions
are imposed at the outer boundary.
 
Regarding the rotating domain, which contains the coordinate singularity at
$\xi=0$, we perform an expansion on a Galerkin basis satisfying the regularity
conditions \cite{BM90}. Moreover, the wave operator is also treated implicitly
with a second-order Crank-Nicolson scheme and the coupling terms containing
first-order time derivatives are treated explicitly by making a second-order
extrapolation from known values at previous time steps.  An implicit treatment
of those terms could be more appropriate but not necessary.

\section{Tests}
\label{s:tests}

We perform two tests of the scheme previously described.

\subsection{Comparison with an analytical solution}
\begin{figure}
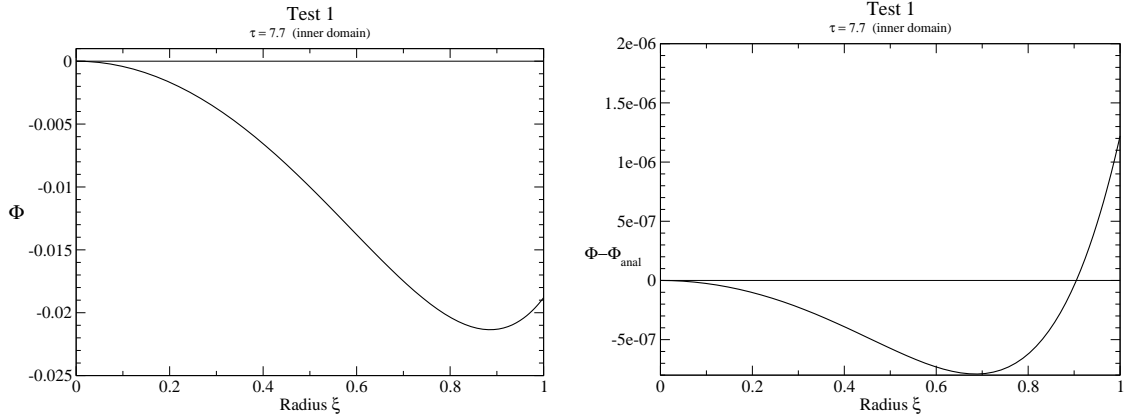

  \centerline{\includegraphics[width=0.55\textwidth]{fig3a.eps}\hspace{1em} 
    \includegraphics[width=0.55\textwidth]{fig3b.eps}}
  \caption[]{
    \label{f:Figure4}  
    Comparison between the numerical and analytical solutions of Test 1. Left
    figure shows the solution at the end of the run $(\tau=7.7)$ and right
    figure the absolute difference between the numerical and the analytical
    solutions in the inner region.}
\end{figure}

The first example deals with the comparison between the numerical and the
analytical solutions, in the restricted case of a helically symmetric operator
obtained from Eq. (\ref{e:rotating_harm_cos_wave_equation}) or Eq.
(\ref{e:rotating_harm_sin_wave_equation}) by setting to zero the time
derivatives. The resulting differential operator, $L$, has the following form
\be\label{ellipt_operator}
L=\frac{d^2}{d \xi^2}+\frac{2}{\xi} \frac{d}{d \xi} 
-\frac{\ell(\ell+1)}{\xi^2}+\left(\frac{2m}{\pi}\right)^2 .
\ee
A straightforward manner to proceed consists in obtaining the source $n$ by
the application of the operator $L$ to a potential $\Phi$
given analytically.  We choose the form of the potential $\Phi$ to be
\be\label{Potential} 
\Phi(\xi)=\frac{\xi^4} {\sqrt \xi} J_{\ell+1/2} (\xi),
\ee 
with $J_{\ell+1/2}$ being the semi-integer Bessel function of the first kind.
In this case, the only modes in (\ref{ellipt_operator}) correspond to $\ell=2$
and $m=2$. The source $n$ computed in this way determines the right-hand-side
in Eqs.~(\ref{e:rotating_harm_cos_wave_equation}) and
(\ref{e:rotating_harm_sin_wave_equation}). The numerical solution obtained by
the application of the present scheme is then compared to the corresponding
analytical solution. The latter is expressed, in the inner domain, as a linear
combination of the exact particular solution (\ref{Potential}) and the regular
homogeneous solution of the operator (\ref{ellipt_operator}), which can be
expressed in terms of the semi-integer Bessel function of the first kind $
\eta (\xi)=\frac{1}{\sqrt \xi} \, J_{\ell+1/2} \, (\xi)$, that can be easily
evaluated because it reduces to the sum of $\ell+1$ terms.  Regarding the
homogenous solution in the external domain, this is expressed in terms of a
second order Bessel function $H_{\ell+1/2}(\xi)$. Once the matching conditions
at $R_*$ and the Sommerfeld boundary conditions at $\xi_D$ are imposed, the
explicit comparison with the analytical solution can be done, finding a very
good agreement (cf. Figure~\ref{f:Figure4} for a quantitative assessment).

The tests above concern the case $\ell=2$ and $m=2$, when a helical symmetry
is present in the source. The number of points used are $N_{\varphi} =
N_{\theta} = 4$, $N_r^{\rm inner} = 65$ in the inner domain (for $\xi\in
[0,1]$), $N_r^{\rm outer} =257$ in the outer domain (for $\xi\in [1, 21]$), and
a time step $\delta \tau = (1024)^{-1}$. This case $\ell=2$ and $m=2$ has been
chosen because it is relevant for gravitational wave emission in General
Relativity.  The choice of a symmetry in the source simplifies the comparison
between analytical and numerical results.

\begin{figure}
  \centerline{\includegraphics[width=0.6\textwidth]{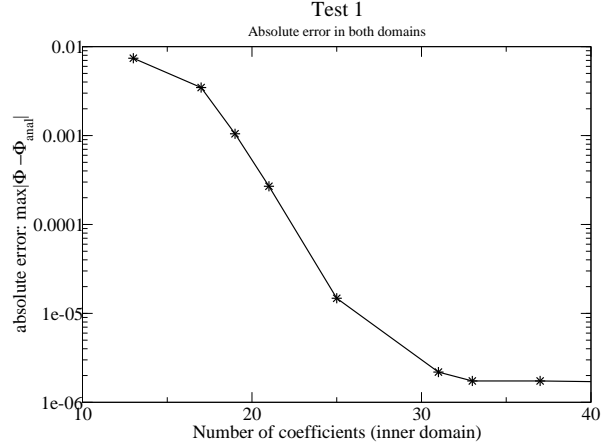}}
  \caption[]{
    \label{f:Figure4bis} 
    Behavior of the accuracy obtained in Test~1 (see also
    Figure~\ref{f:Figure4}) as a function of the number of radial
    spectral coefficients. The displayed numbers correspond to the
    inner domain; in the outer domain eight times more coefficients have been
    used. 
}
\end{figure}
We have studied the convergence behavior of the code with the same settings,
but changing the number of radial coefficients in both domains, while keeping
fixed the ratio $N_r^{\rm outer} / N_r^{\rm inner} = 8$. In particular, we
have checked that the difference between the numerical and analytical
solutions would decay as $e^{-N_r}$, which is the expected rate for spectral
approximation series of smooth functions. This check is displayed in
Figure~\ref{f:Figure4bis}, where the discrepancy saturates at about $10^{-6}$
level, due to numerical limitations. There are two sources accounting for this
accuracy level, whose behaviors were clearly identified in numerical
investigations: first, the time finite-differencing errors, which are linked
to the second-order time-scheme, and then the use of approximate boundary
conditions (we remind that Sommerfeld boundary conditions are exact only for
$\ell =0$ modes) at $\xi = \xi_D$. We have checked that these latter errors
would decay as $\xi_D^{-2}$, which is the predicted rate (see
also~\cite{NovakB}).

\subsection{Conservation of helical symmetry}
The second test to our resolution scheme studies how a solution $\Phi$ to
Eqs.~(\ref{e:wave_equation}) and (\ref{e:rotating_wave_equation}) ``tracks'' a 
symmetry present in the prescribed matter source.  More specifically, we
consider a source $n$ which is rigidly rotating with respect to the inertial
frame with an angular velocity $\Omega$. We choose its form as
\be\label{source_gal_test}
n=n_0 \left[
\left( \frac{r}{R_*} \right) ^4-\left (\frac{r}{R_*} \right)^2 \right] 
\sin ^2\theta \cos(2\varphi
+2\Omega t).
\ee
This source is symmetric with respect to the helical vector $\zeta \equiv
\partial_t-\Omega
\partial \varphi $, i.e.  
\be\label{Killing_vec_source} 
\left(
  \frac{\partial}{\partial t} -\Omega \frac{\partial}{\partial \varphi}
\right) \,n=0 .  
\ee 
Because of the existence of this symmetry, we expect the solution $\Phi$ to
present also the helical symmetry once a sufficient time has been elapsed (once
it has reached some ``stationary'' regime).

We apply to this simple case the scheme described in section
\ref{numerical_strategy}.  Regarding the inner rotating domain, since the
only non-vanishing terms correspond to $\ell =2$ and $m=2$,
Eqs.~(\ref{e:rotating_harm_cos_wave_equation}) and
(\ref{e:rotating_harm_sin_wave_equation}) read as follows:
\begin{eqnarray}
&&\frac{\partial^2 a^< }{\partial \tau^2 } 
+ (2 \pi) \left( 4 \frac{\partial b^< }
{\partial \tau} -(2\pi)4 a^< \right)    \nonumber  \\ 
&& -C^2 \left( \frac{\partial^2 }{\partial \xi^2} +\frac{2}{\xi}
\frac{\partial}{\partial \xi} -\frac{6}{\xi^2} \right) a^<= \bar{n}(\xi) 
\label{e:rotating_teat_a}\\
&& \frac{\partial^2}{\partial \tau^2} b^< -(2 \pi) \left( 4  \frac{\partial
  a^< }{\partial \tau}+4 (2\pi)b^< \right) \nonumber \\
&& -C^2 \left( \frac{\partial^2}
 {\partial_\xi^2}+\frac{2}{\xi} \frac{\partial}
 {\partial \xi}
-\frac{6}{\xi^2} \right) b^< =0
\label{e:rotating_trast_b}
\end{eqnarray}
where for shortness, we have written respectively $a^<=a^<_{22} $, $b
^<=b^<_{22} $ and $\bar{n}=R_*^2\cdot C^2 n_{22} $.  
Note that in the inner
rotating frame, the source $\bar{n} $ does not depends on $\tau$ and
$\bar{n}^s=0$.  In the exterior non-rotating domain we solve the similar
expressions coming from Eqs.~(\ref{e:galil_harm_cos_wave_equation}) and
(\ref{e:galil_harm_sin_wave_equation}).

Initial conditions are chosen such that $a=b=0$ at $\tau = \tau_0$, and the
evolution is induced by the source $\bar{n}$ (see Fig.~\ref{f:Figure6}a, 
where
the profile at the first time-step is given).  This was computed in a such way
to ensure the continuity of the solution and its derivative across the
boundary $r=R_*$.  The dynamical evolution began at time $\tau =
\tau_\mathrm{o}=0$, and the numerical settings were slightly different from
those of the first tests, namely: $N_r^{\rm inner}=33, N_r^{\rm outer} = 129,
N_\theta = 4, N_\varphi = 4$ and $\delta \tau = (512)^{-1}$.

\begin{figure}
  \centerline{\includegraphics[width=0.6\textwidth]{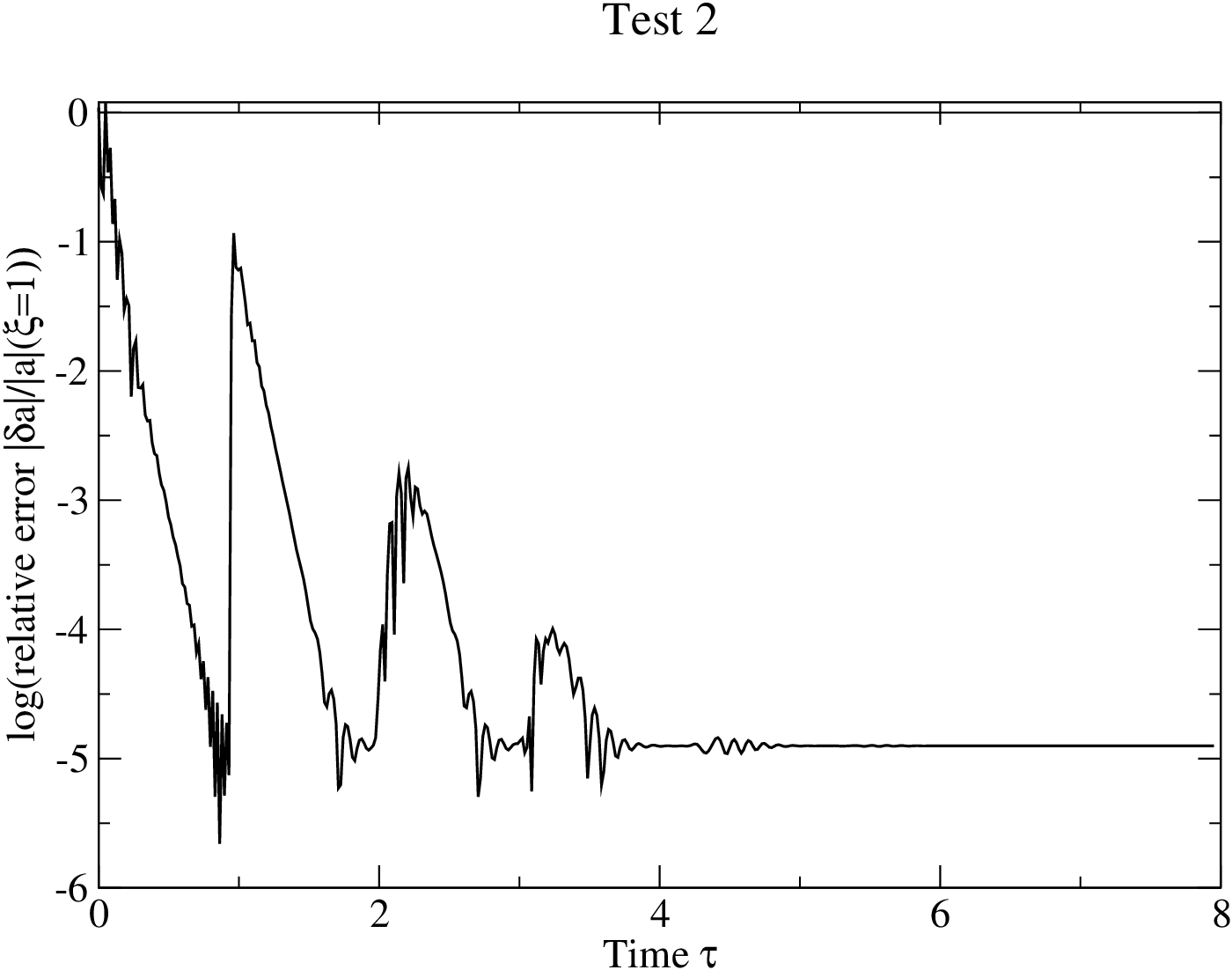}}\vspace{1em}

  \centerline{\includegraphics[width=0.55\textwidth]{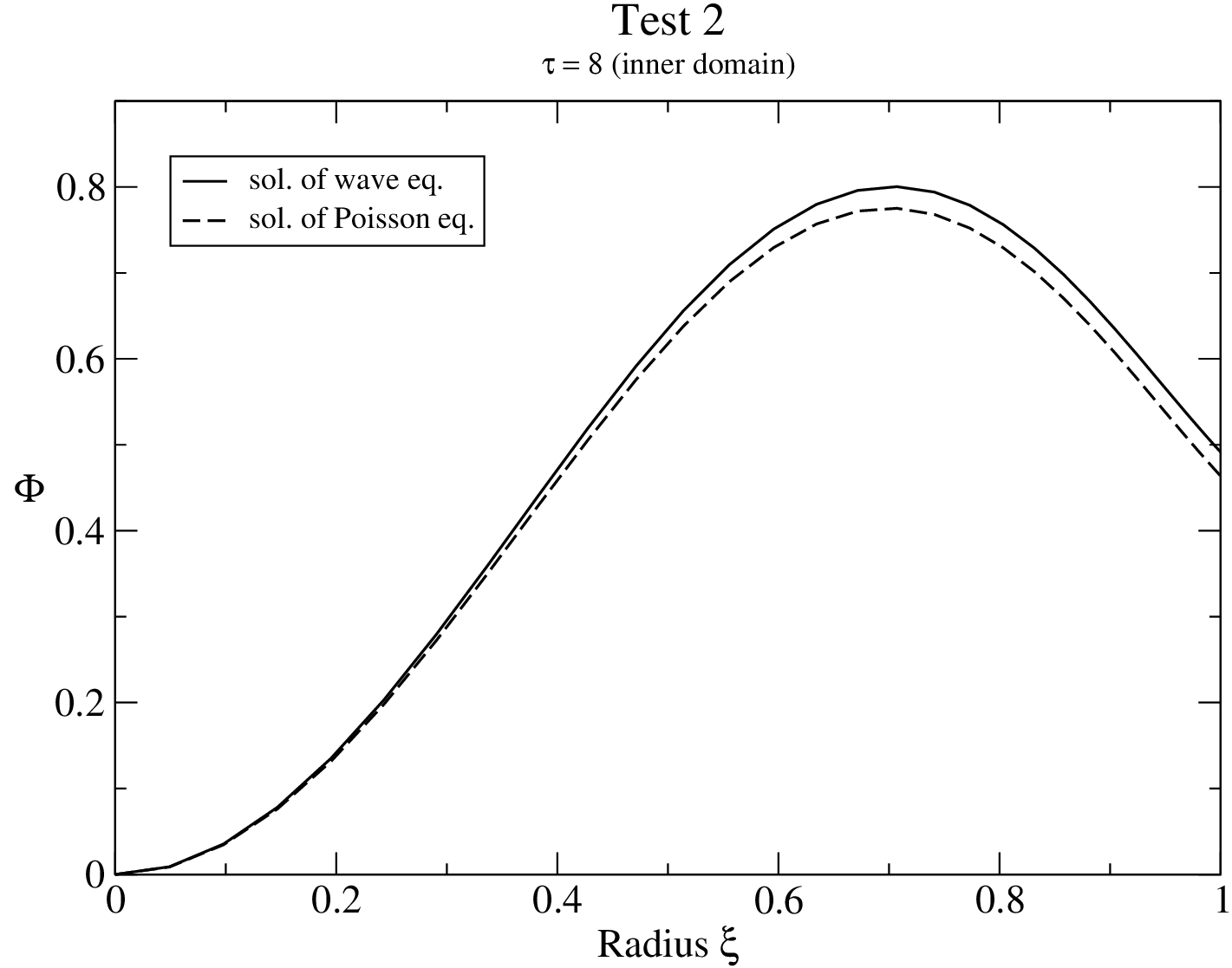}
    \includegraphics[width=0.55\textwidth]{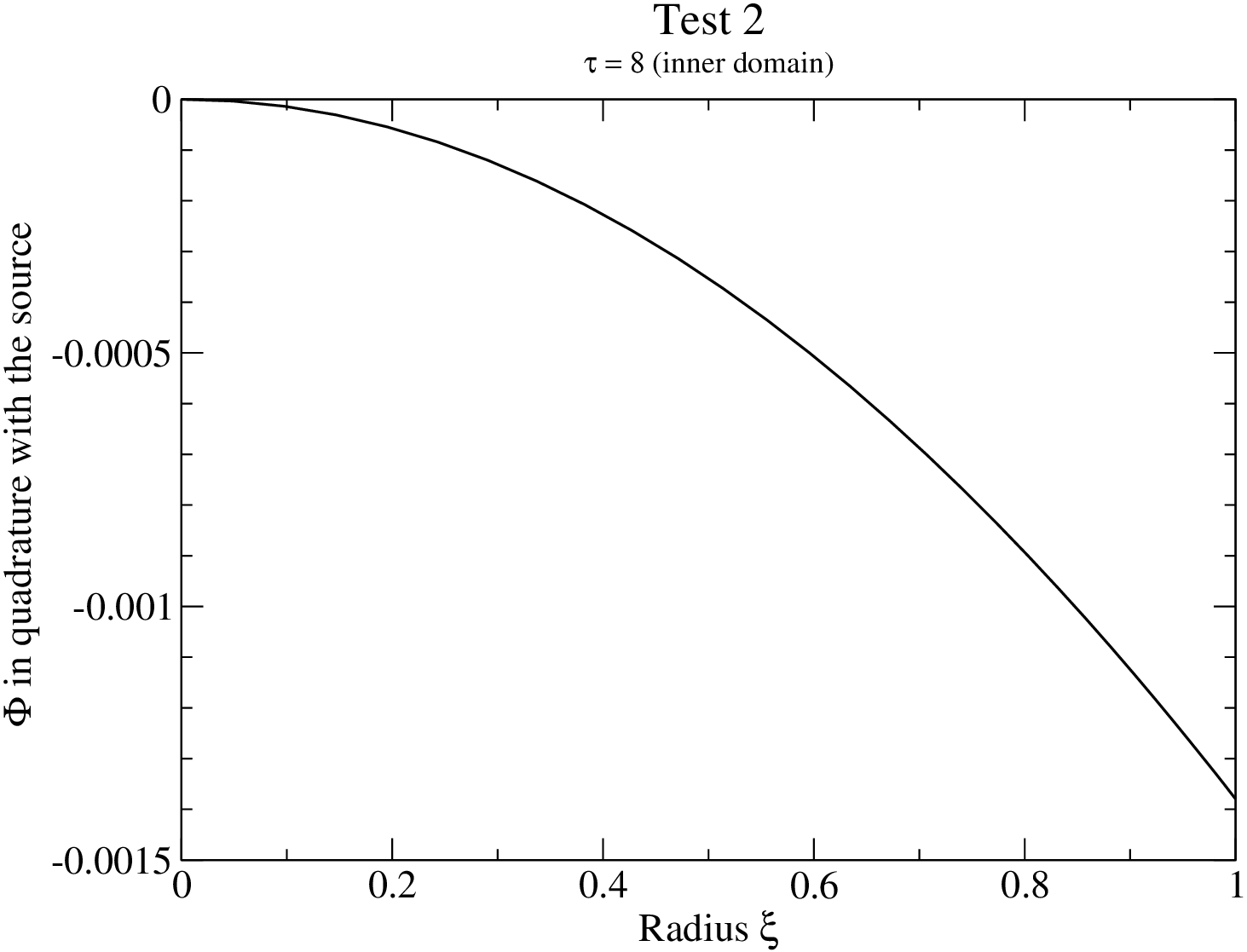}}
  \caption[]
  {
    \label{f:Figure3}  
    Fig. \ref{f:Figure3}a: Relative error $\delta a$ for Test 2 in a 
    logarithmic
    scale versus the time $\tau$, for $R_L/R_*=3$. The huge spike at time
    $\tau=1$ is due to terms containing first time derivatives in
    Eqs.~(\ref{e:rotating_harm_cos_wave_equation}),
    (\ref{e:rotating_harm_sin_wave_equation}) that are switched on only at
    $\tau =1$ in order to avoid a strong oscillation at the beginning of the
    simulation. The plateau reached by the error is due to
    time-differentiation and scales like $\delta \tau^2$.
    Fig.~\ref{f:Figure3}b: The actual solution of the wave equation in the
    rotating region (upper curve) and the solution of the Poisson equation.
    Fig.~\ref{f:Figure3}c: Component of the solution $\Phi$ in quadrature
    with the source.}
\end{figure}

\begin{figure}
\centerline{  \includegraphics[angle=-90, width=0.6\textwidth]{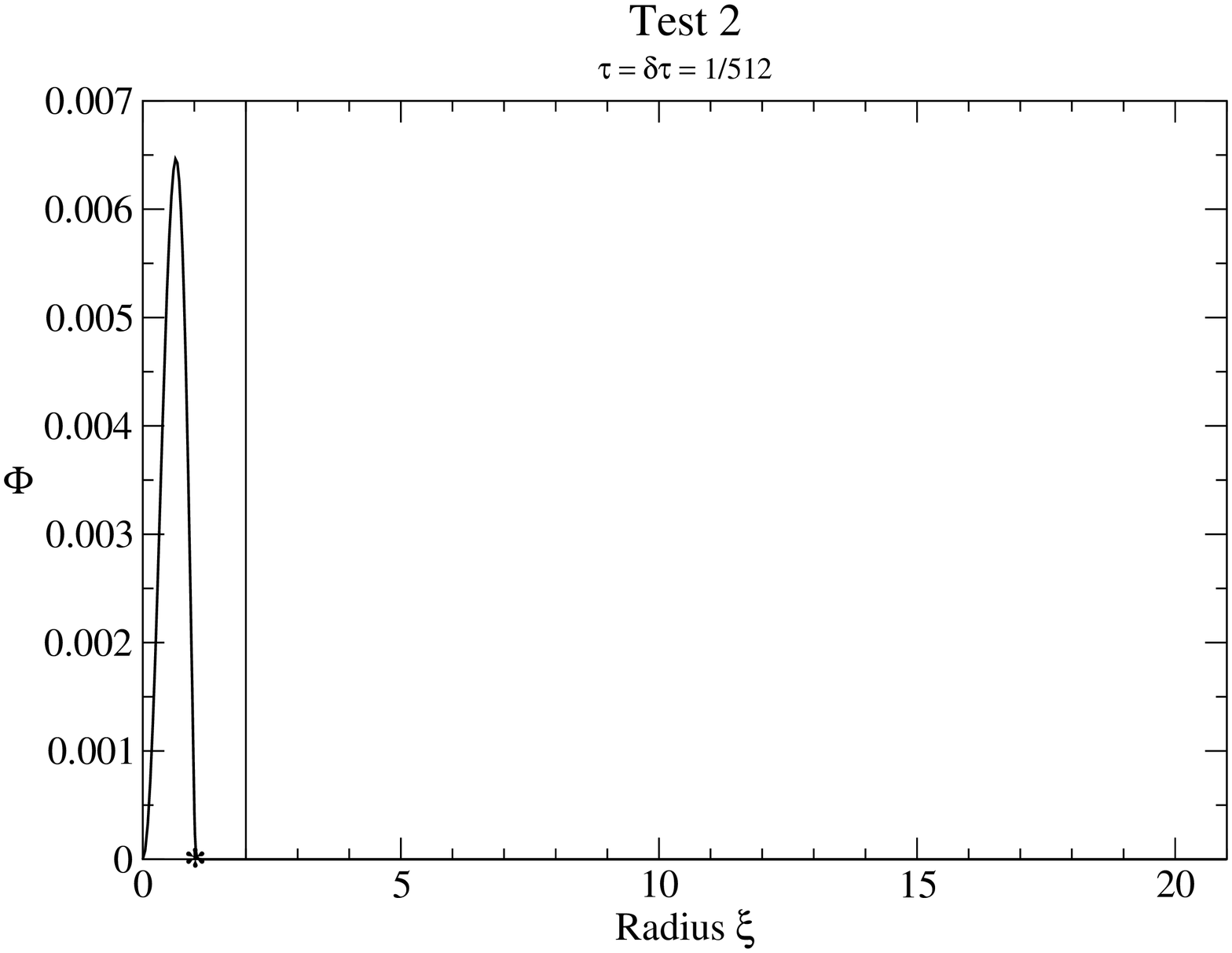}
  \includegraphics[angle=-90, width=0.6\textwidth]{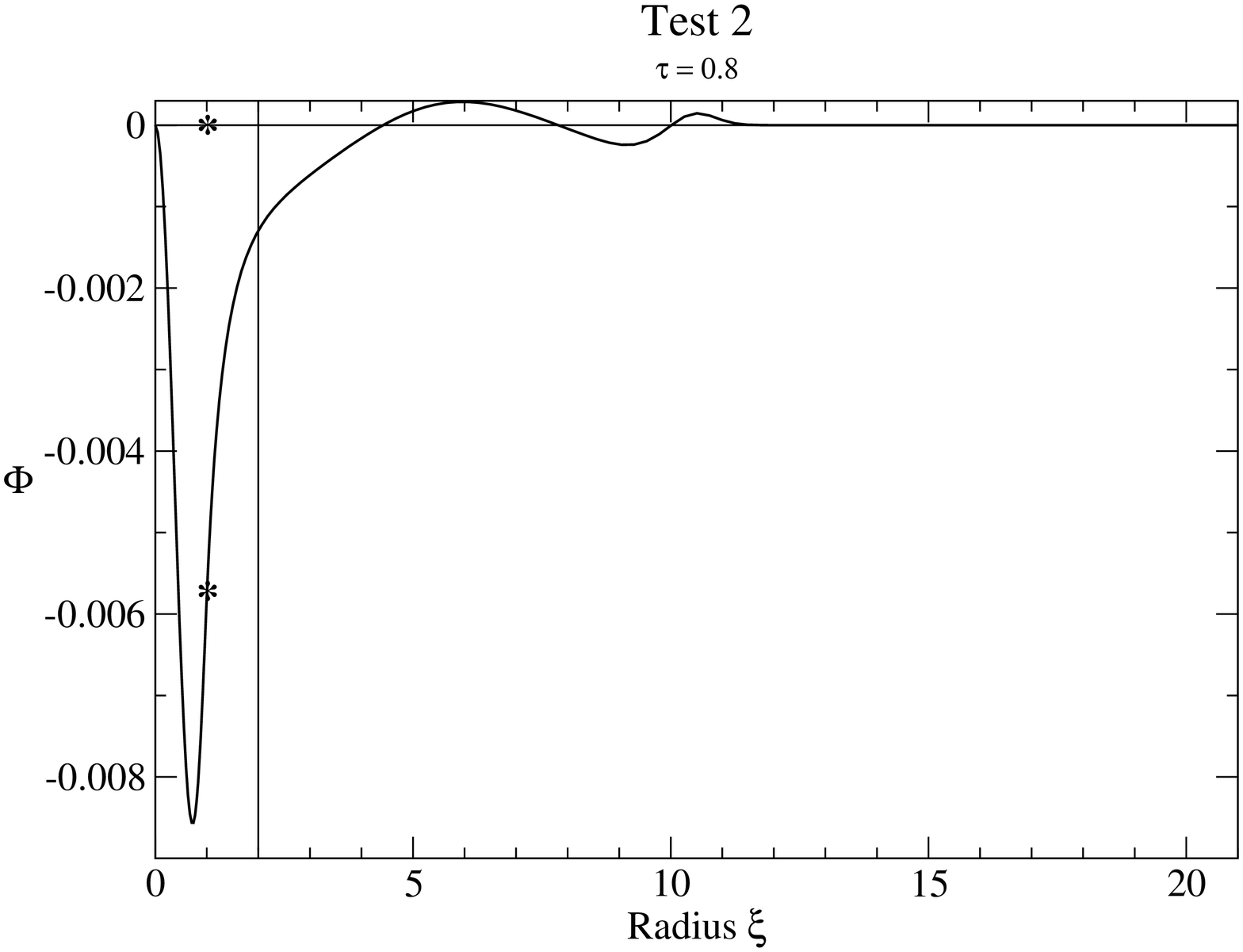} }
\centerline{  \includegraphics[angle=-90, width=0.6\textwidth]{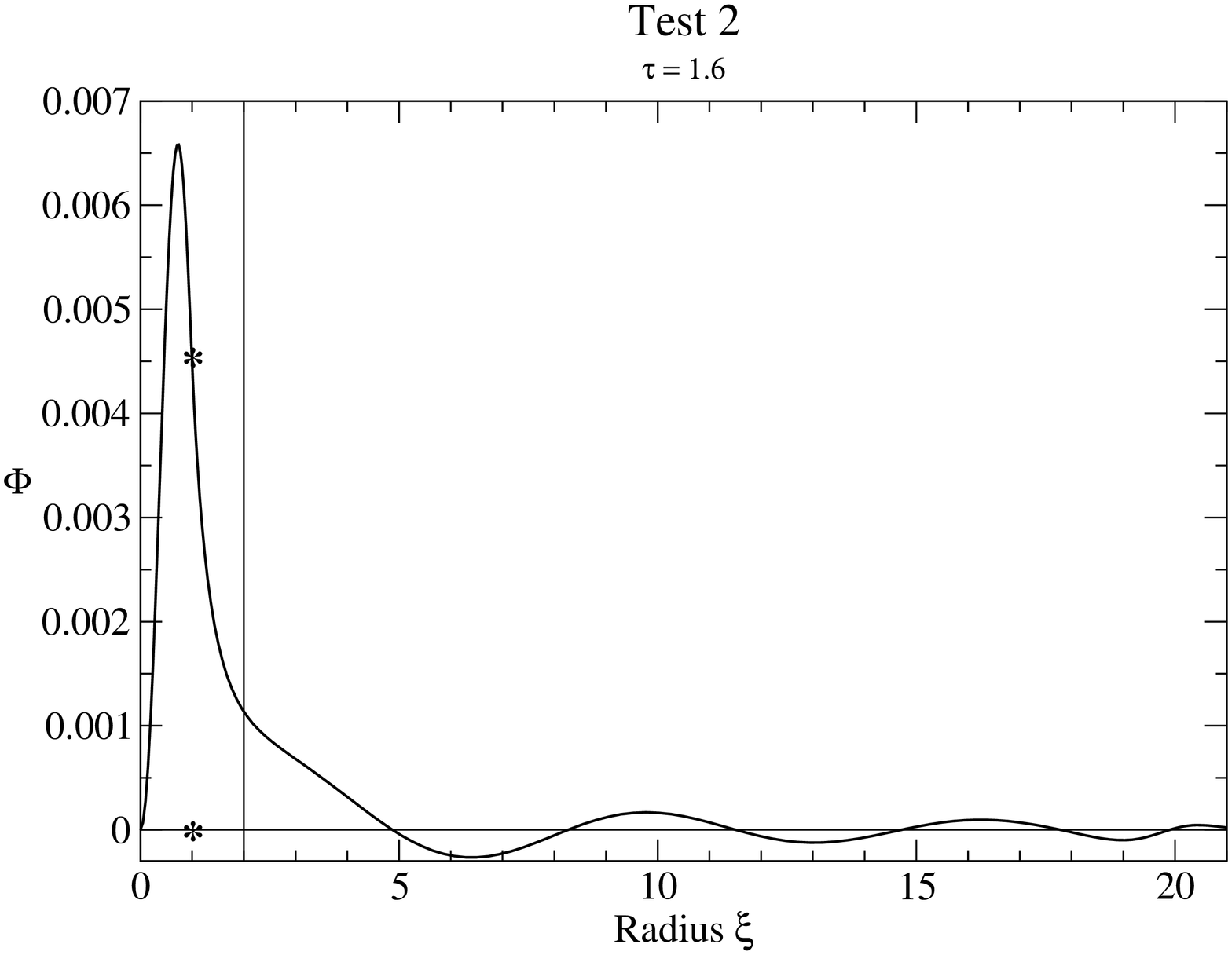} 
  \includegraphics[angle=-90, width=0.6\textwidth]{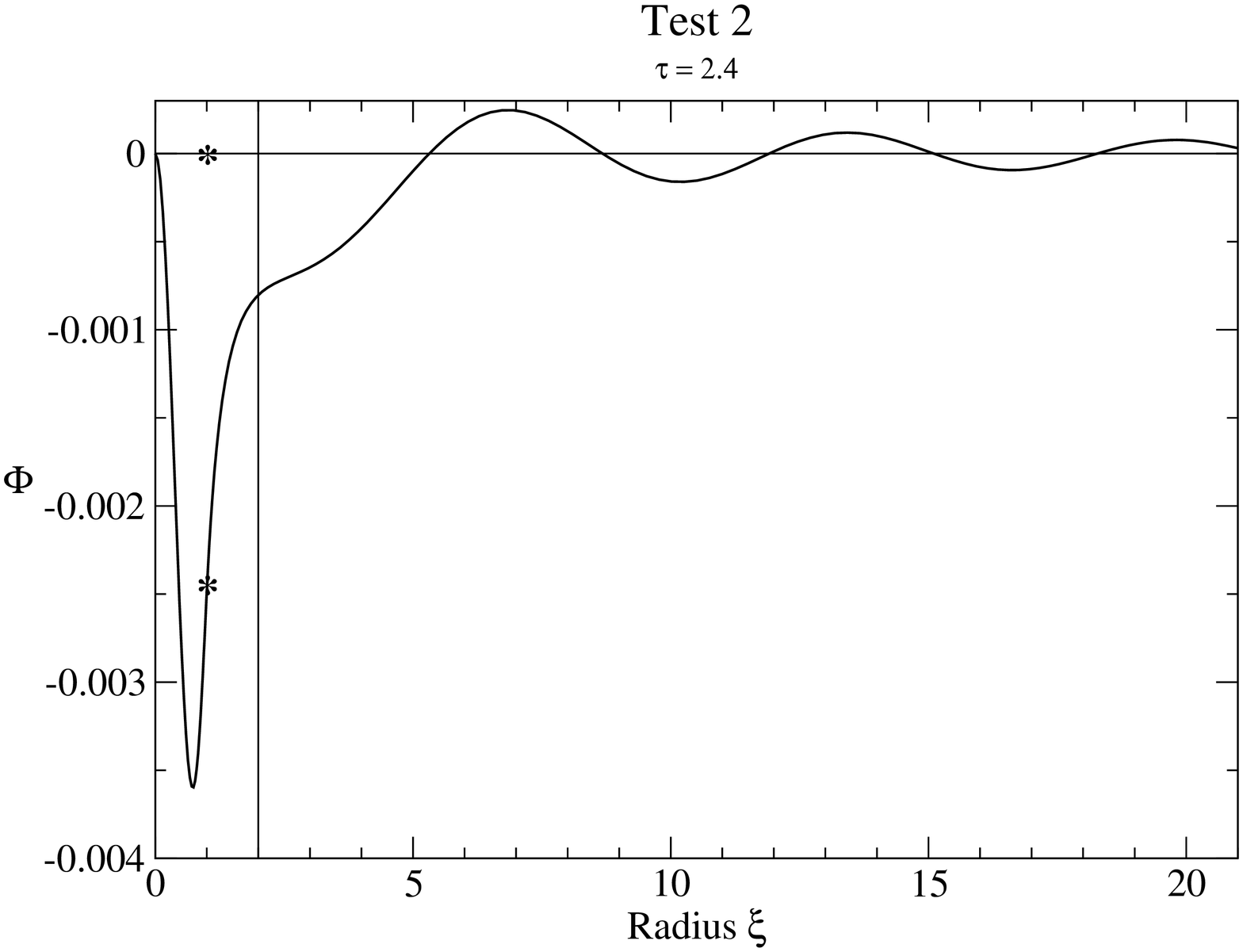} }
\centerline{  \includegraphics[angle=-90, width=0.6\textwidth]{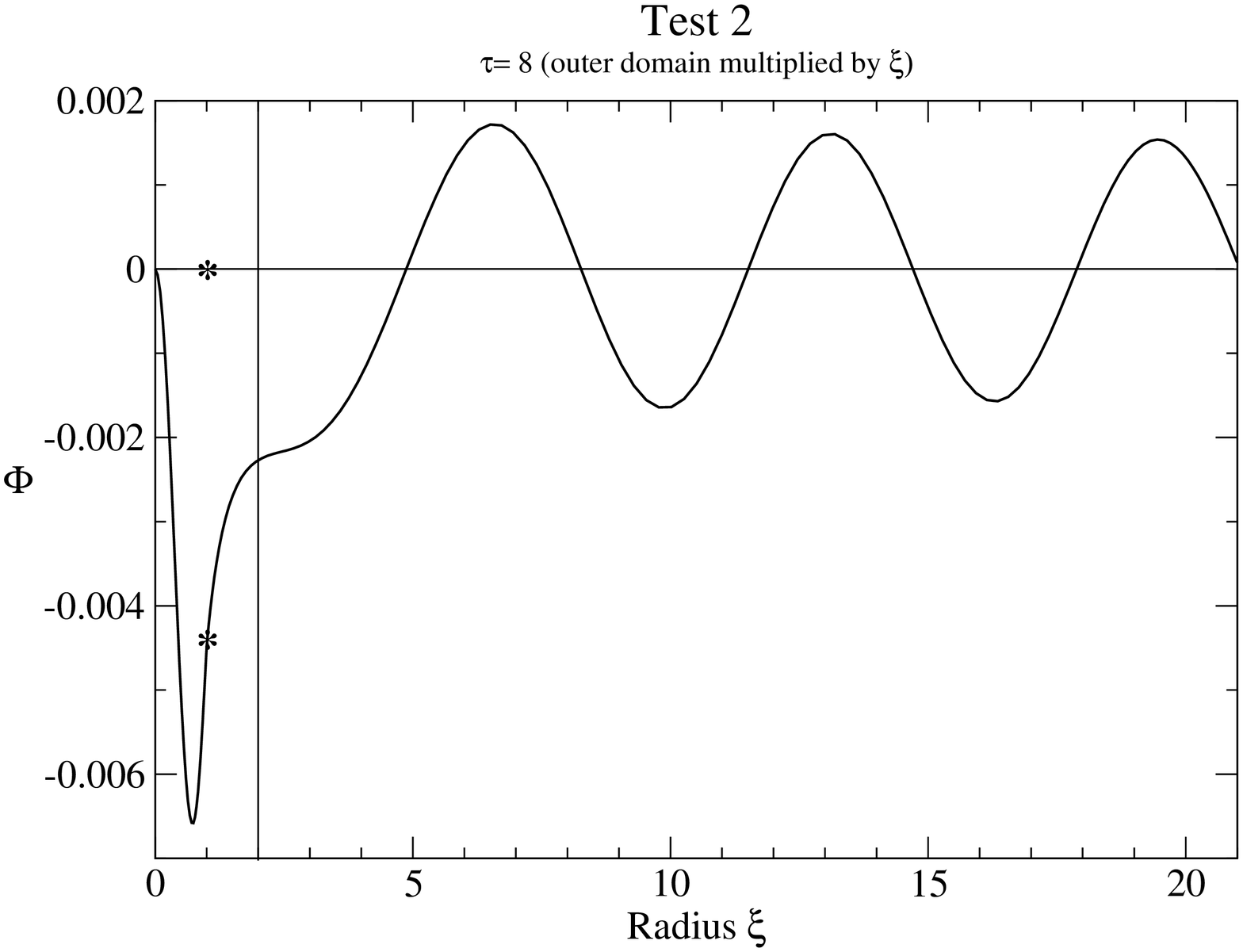}
  \includegraphics[angle=-90, width=0.6\textwidth]{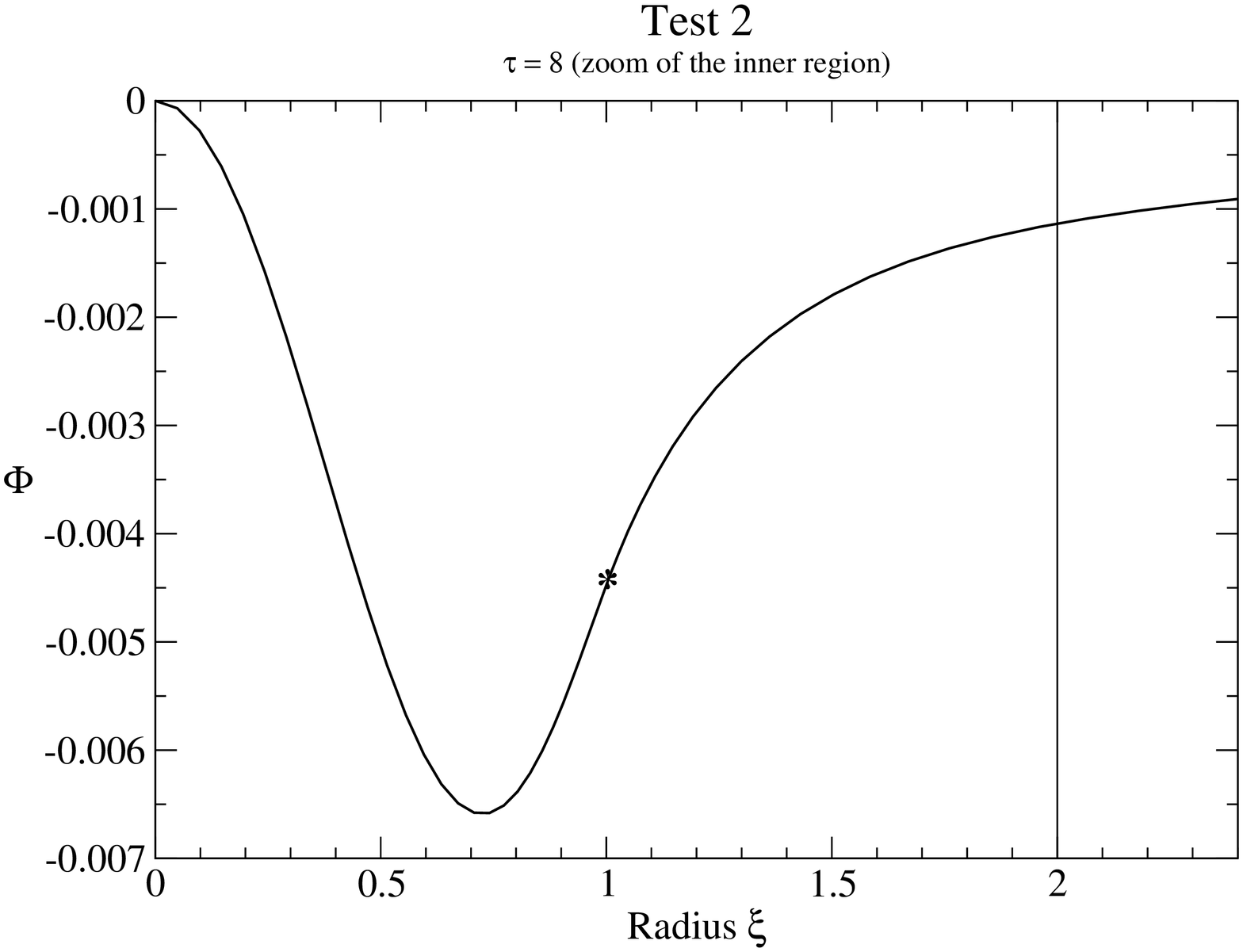}}
  \caption[]{
    \label{f:Figure6}
    Figures \ref{f:Figure6}.a-e show some snapshots of the solution in Test 2,
    with $R_L/R_*=2$ at different times. At $\xi=1$ the ``star'' shows the
    frontier between the rotating and inertial domains. The vertical line at
    $\xi=2$ indicates the light cylinder. On Fig.~\ref{f:Figure6}.e the
    solution in the external region was multiplied by $\xi$ in order to better
    visualize the solution at the edge of the external grid.
    Figure~\ref{f:Figure6}.f is a zoom of the solution inside and near the
    light cylinder. The dimensionless ``light velocity'' is $C=4\pi$.}
\end{figure}

We study the emergence of the helical symmetry in the evolving dynamical
solution in two different manners:

\noindent {\it i\/}) First we assess the verification of the symmetry
conditions given by the Eq.~(\ref{Killing_vec_source}), for the case of the
constructed numerical solution.  This provides a good self-consistent test of
the accuracy of the solution.  In the present case, we compute the quantity
$\delta a$, defined as
\be\label{ad_Killing_vec_solution} 
\delta a:=\frac{\partial a} {\partial \tau} -\Omega 
\frac{\partial a}{\partial \varphi}
\ee
where $a$ is considered as a function of $\xi$ and $\tau$, and we study its
evolution in time. Figure \ref{f:Figure3}.a shows the evaluation of
(\ref{ad_Killing_vec_solution}) on the numerical solution: after the transient
due to the switch on of the evolution, the solution must asymptotically verify
$\delta a = 0$.  Relative error, computed at $\xi=1$ in the outer region, is
plotted versus the time $\tau$.  Figure \ref{f:Figure3}.b shows the difference
between the computed solution and the solution of the Poisson equation for
$\Phi$ (instead of the wave equation~(\ref{e:wave_equation}), to show
propagation effects). The actual solution differs qualitatively with respect
to the non-propagating one by a phase-($\pi/2$) quadrature term that is
showed in Figure \ref{f:Figure3}.c.  This term is responsible of the
radiation reaction force (or braking force), and we have checked that this
term scales as $(R_*/r)^5$ as predicted by the analytical properties for an
$\ell=2$ propagation mode. Note that the solution of the Poisson equation, i.e.
satisfying the equation $\bar{\Delta} a^<= -\bar{n}$, is 
\be
\label{solnew} 
a^<(\xi,0)=n_0 \left( \frac{\xi^4}{14}-\frac{\xi^6}{36}+\alpha \xi^2 \right),
\quad a^>(\xi,0)= \frac{n_0\beta}{\xi^3}, 
\ee 
where $\alpha=(1/20-1/14)$ and $\beta=(1/45-1/35)$.

Finally, Figure~\ref{f:Figure6} presents some radial profiles at various
moments of the numerical solution to Eqs.~(\ref{e:rotating_teat_a}) and
(\ref{e:rotating_trast_b}).

\begin{figure}
  \centerline{\includegraphics[angle=-90, width=0.6\textwidth]{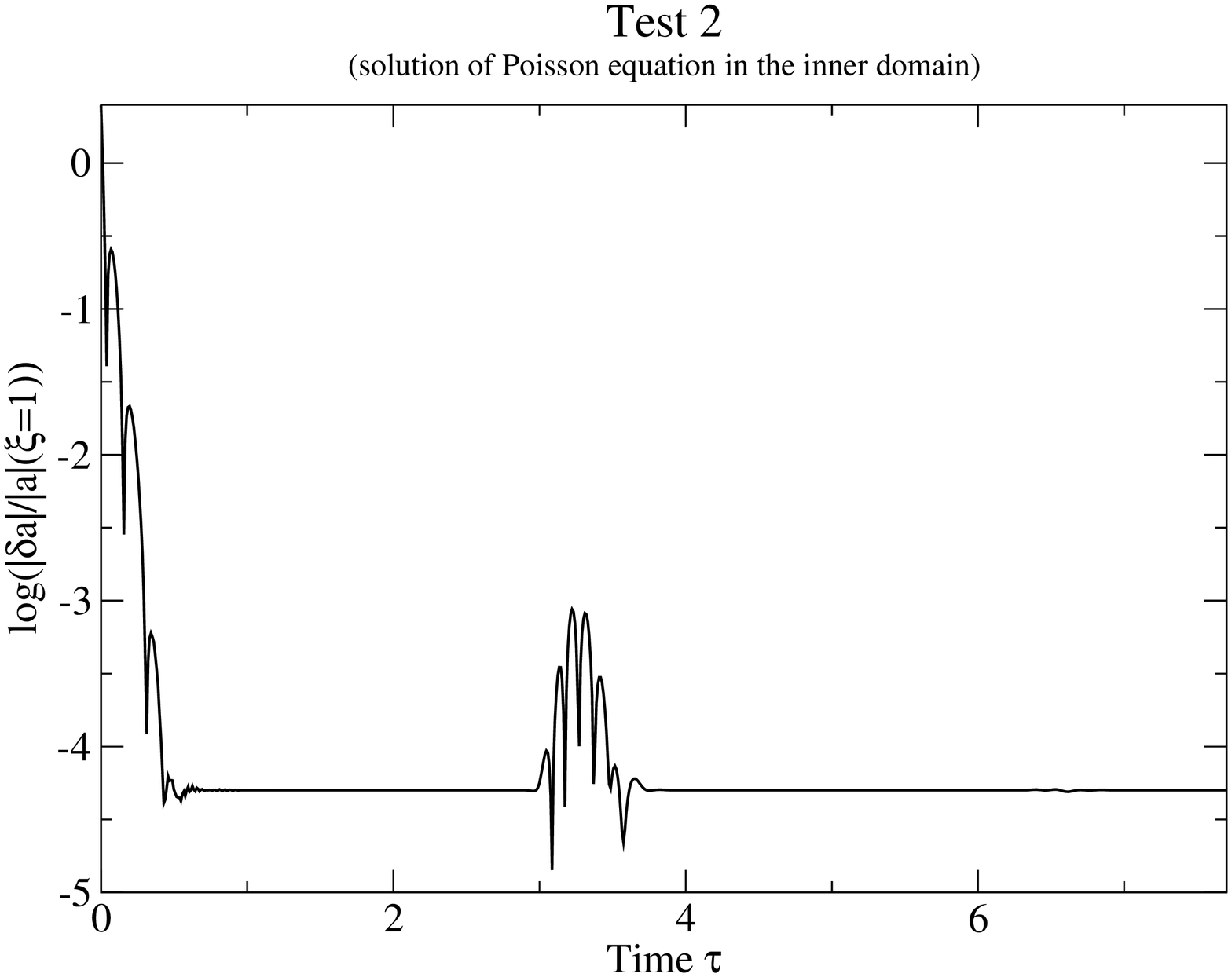}}
  \caption[]{
    \label{f:Figure5}  
    Test of the helical symmetry $\delta a$ as in Fig.~\ref{f:Figure3}a.}
\end{figure}

\noindent {\it ii\/}) A second way to proceed in order to study the emergence
of the helical symmetry in the solution, consists in comparing the previous
dynamical solution with the one corresponding to a system obtained by
neglecting the time derivative terms in
Eqs.~(\ref{e:rotating_harm_cos_wave_equation})-(\ref{e:rotating_harm_sin_wave_equation}),
in the inner domain.  In this case the problem in the rotating region becomes
an elliptic one, and it can be solved easily by inverting the operator $L$
given in Eq. (\ref{ellipt_operator}), whereas we still solve for the wave
operator in the outer domain.  Note that it is crucial for the ellipticity of
this operator $L$ the fact that the rotating grid is set inside the light
cylinder. If this were not be the case, we would have a mixed-type
differential operator (see e.g. \cite{Torre03}).  Given the second-order
Crank-Nicolson scheme we have employed, results computed with the method {\it
  i\/}) and the method {\it ii\/}) differ only by quantities of second order
in $(\delta \tau)^2$.  As a complementary test to the emergence of the helical
symmetry, Figure~\ref{f:Figure5} shows evaluation of
(\ref{ad_Killing_vec_solution}) on the numerical solution to the
symmetry-reduced operator~(\ref{ellipt_operator}). Note that the solution
reaches the stationary regime much faster than when solving the full wave
equation. The big bump appearing at the time $\tau \sim 3.2$ and the small one
at the time $\tau \sim 5.5$ are generated by a partial reflexion at the
external boundary of the external domain due to the approximate outgoing
Sommerfeld boundary conditions [which are exact for $\ell=0$; see
Eq.~(\ref{Somme})]. We have checked that the amplitude of the bumps scales as the
inverse of the square of the dimension of the external grid, as predicted
by the theory.

\section{Discussion and conclusions}
\label{s:conc}

The rationale of this work is the following: a class of physical problems
(rotating stars or orbiting stars/black holes) are easily treated in corotating
grids. However, analytical and numerical problems show up associated with the
existence of a light cylinder. We have described a technique to match the
solution computed in an inner rotating domain of external radius $R_*$, with
the solution in an outer non-rotating domain under the condition $R_* < R_L$,
where $R_L=c/ \, \Omega$ is the light cylinder radius.  The matching technique
is independent of the manner of obtaining the solutions in each domain.  In
typical physical situations, the restriction $R_* < R_L$ on the radius is not
severe: for example in the worst case for orbiting neutron stars, we have
$\Omega \sim 4000 \;s^{-1}$ \cite{Bejge05} and $R_* \sim 20 $ km, providing a
ratio $R_L/R_*\sim 4$. As a more extreme case, this ratio for orbiting black
holes before the merger is not smaller than 2.  A typical problem to be
treated with the technique described here, is the case two orbiting NS, for
which the parameters of the orbit are known, and for which we want study small
amplitude motions inside each star (convection r-modes and/or elliptical
cylinder instabilities).  The numerical study of these phenomena is easier in
a domain in which the surface of the star does not move and boundary condition
can be exactly imposed, as compared with a star traveling through the grid.

Regarding the numerical implementation, this is not the only scheme making use
of multi-domain spectral methods in order to deal with the resolution of
partial differential equations in rotating domains (cf. \cite{ScheePLKRT06} in
the general dynamical case and \cite{LauP07} in the helically symmetric
situation).  The specificity of our approach is the use of a rotating domain
(our inner domain) that, unlike the case in other schemes, does not extend
up to infinity. In particular our inner rotating domain does not go beyond the
light cylinder, thus providing a simple treatment of this analytical and
numerical issue.  The results we have obtained are accurate, and quantities
like the back-reaction force are correctly computed.  We conclude that our
``stroboscopic'' matching technique is a simple and efficient tool for the
description of relativistic rotating systems, and in particular for avoiding
analytical and numerical problems linked with the existence of a light
cylinder.

The technique described can be extended, {\it mutata mutandis}, to vectorial
or tensorial equations, in particular choosing some convenient triad (or tetrad
depending on the formulation), and projecting onto it the tensor components in
order to deal with scalar quantities.

Finally, the case involving a non-constant angular velocity can also be
treated: if the variation of the rotation period $P$ in time is known, we can
perform the coordinate transformation $t=t(\tau)$ generalizing
Eqs.~(\ref{xi_tau}), in such a way that wave equations correcting
Eqs.~(\ref{e:ad_wave_equation}) and (\ref{e:ad_rotating_wave equation}) 
with extra terms in $\frac{\partial
  \,t}{\partial \, \tau} $ are obtained.  These additional terms are harmless
for our scheme, provided that the ratio $R_L/R_*$ is sufficiently large.  All
those possibilities are yet to be explored.  
\vspace{0.2in}

\noindent
{\bf Acknowledgments}
\vspace{0.1in}

\noindent
We thank \'Eric Gourgoulhon and Philippe Grandcl\'ement for helpful
discussions.  SB and JN were supported by the A.N.R. Grant 06-2-134423
entitled ``Mathematical methods in general relativity'' (MATH-GR).
JLJ acknowledges the support of the Marie
Curie European Reintegration contract MERG-CT-2006-043501 
within the 6th European Community Framework Program, and the
hospitality of the Observatoire de Paris. 
\vspace{0.2in}

\noindent
{\bf References}
\vspace{0.1in}

\bibliographystyle{prsty}

\end{document}